\documentclass[useAMS,usenatbib]{mn2e} 
\usepackage{graphicx}
\usepackage{epsfig}
\usepackage{amsmath}

\newcommand{\hs}{\hspace{1mm}} 
\newcommand{\apj}{ApJ}
\newcommand{\aap}{A\&A} 
\newcommand{\apjl}{ApJL}
\newcommand{\mnras}{MNRAS} 
\newcommand{\aj}{AJ}
\newcommand{\apjs}{ApJS} 
\newcommand{\nat}{{\it Nature}}
\newcommand{\araa}{ARA\&A}

\newcommand{\lya}{Ly$\alpha$}

\def\plottwo#1#2{\centering \leavevmode \epsfxsize=.45\textwidth
\epsfbox{#1} \hfil \epsfxsize=.45\textwidth \epsfbox{#2}}

\def\lsim{~\rlap{$<$}{\lower 1.0ex\hbox{$\sim$}}}

\def\gsim{~\rlap{$>$}{\lower 1.0ex\hbox{$\sim$}}}

\title[\lya{} EW vs. SFR(\lya{})/SFR(UV)]{Star Formation
Indicators and Line Equivalent Width in Ly$\alpha$ Galaxies}

\author[M.~Dijkstra \& E.~Westra]{Mark
  Dijkstra$^1$\thanks{E-mail:mdijkstr@cfa.harvard.edu} and Eduard
  Westra$^2$\\
  $^{1}$Astronomy Department, Harvard University, 60 Garden Street,
  Cambridge, MA 02138, USA\\
  $^{2}$Smithsonian Astrophysical Observatory, 60 Garden Street,
  Cambridge, MA 02138, USA}

\begin{document}

\date{\today} \pagerange{\pageref{firstpage}--\pageref{lastpage}}
\pubyear{2009}

\maketitle

\label{firstpage}
\begin{abstract}
  The equivalent width (EW) of the \lya{} line is directly related to
  the ratio of star formation rates determined from Ly$\alpha$ flux
  and UV flux density [SFR(\lya{})/SFR(UV)]. We use published data
  --in the literature EW and SFR(\lya{})/SFR(UV) are treated as
  independent quantities-- to show that the predicted relation holds
  for the vast majority of observed \lya{} emitting galaxies
  (LAEs). We show that the relation between EW and SFR(\lya{})/SFR(UV)
  applies irrespective of a galaxy's ``true'' underlying star
  formation rate, and that its only source of scatter is the variation
  in the spectral slope of the UV continuum between individual
  galaxies. The derived relation, when combined with the observed
  EW distribution, implies that the ratio SFR(UV)/SFR(Ly$\alpha$) is
  described well by a log-normal distribution with a standard
  deviation of $\sim 0.3-0.35$. This result is useful when modelling
  the statistical properties of LAEs. We further discuss why the
  relation between EW and SFR(\lya{})/SFR(UV) may help identifying
  galaxies with unusual stellar populations.
\end{abstract}

 
\section{Introduction}
\label{sec:intro}
Lyman $\alpha$ (hereafter Ly$\alpha$) equivalent width represents a
fundamental quantity in the study of Ly$\alpha$ emitting galaxies.
The equivalent width (EW) of the \lya{} line emitted by galaxies is a
sensitive indicator of the initial mass function (IMF) or gas
metallicity from which stars form \citep[e.g.][]{S02,S03}. The
existence of large equivalent width (rest-frame $\mathrm{EW} \ga
240$\,\AA{}) \lya{} emitters has led to speculation on whether
population III formation from pristine gas may actually have been
observed \citep{MR02,J06,DW07}. However, this interpretation depends
sensitively on the details of radiative transfer through both the
intergalactic medium \citep[IGM, see e.g.][for a discussion]{S08}, and
the interstellar medium \citep[ISM, e.g.][]{N90,H06,Fi}. It is safe to
conclude that at present, no truly convincing candidates for
population III galaxies exist. The search for large equivalent width
\lya{} emitters -- and population III galaxy formation -- is a key
science driver for the observational community, while modelling \lya{}
radiative transfer in large EW emitters is a challenge for theorists.

The main goal of this paper is to draw attention to the fact that
\lya{} EW is frequently discussed independently from the quantity
SFR(\lya{})/SFR(UV)
\citep[e.g.][]{Ajiki03,Fujita03,Venemans04,Shima06,Gronwall07,Tapken07,P08}.
This quantity denotes the ratio of the star formation rates derived
from the observed \lya{} flux and rest-frame UV flux density. One can
consider the quantity SFR(\lya{})/SFR(UV) as an ``alternative''
measurement of EW, because the quantities EW and SFR(\lya{})/SFR(UV)
are directly related (see \S~\ref{sec:basis} of this paper, and
e.g. \citealp{DW07}, \citealp{Rauch08},  \citealp{Dayal08}, \citealp{Nagamine08}, \citealp{Nilsson09}).  Since
EW represents a fundamental property of \lya{} emitting galaxies, a
more detailed investigation of its relation to the ratio
SFR(\lya{})/SFR(UV) is warranted.

We derive the relation between EW and SFR(\lya{})/SFR(UV), and
compare with observations in \S~\ref{sec:result}. We discuss the
implications of our results in \S~\ref{sec:conc}. Throughout this
paper we denote the rest frame equivalent width by REW, and the
observed equivalent width by OEW. The two are related by
REW=OEW$/(1+z)$. When we write ``EW'' this refers to both OEW and REW.

\section{The correlation between EW and SFR(\lya{})/SFR(UV)}
\label{sec:result}
\subsection{The basis of the Correlation}
\label{sec:basis}
The star formation rate derived from the \lya{} line is obtained from
the better-calibrated H$\alpha$ star formation indicator given by
SFR(H$\alpha$)=$7.9\times 10^{-42}L_{{\rm H}\alpha}$\hs$M_{\odot}$
yr$^{-1}$, where $L_{{\rm H}\alpha}$ is the H$\alpha$ luminosity in
erg s$^{-1}$ \citep[e.g.][]{K98}. For case-B recombination the
corresponding \lya{} luminosity of the source is $L_{{\rm
Ly}\alpha}=8.7L_{{\rm H}\alpha}$, and thus SFR(\lya{})$=9.1\times
10^{-43}L_{{\rm Ly}\alpha}$ $M_{\odot}$ yr$^{-1}$.

Similarly, the star formation rate obtained from the UV luminosity
density is generally given by SFR(UV)$=1.4\times 10^{-28}L_{\nu,UV}$
$M_{\odot}$ yr$^{-1}$ \citep{K98}. $L_{\nu,UV}$ is the UV luminosity
density in erg\,s$^{-1}$ Hz$^{-1}$ measured at $\lambda=\lambda_{\rm
UV}$.  To be consistent with published work we shall assume
$\lambda_{\rm UV}=1400$ \AA. Formally, the standard conversion
applies when $\lambda_{\rm UV}=1500-2800$ \AA\hs \citep[e.g.][]{K98},
and the assumed star formation calibrators may not be completely
accurate. This only introduces a minor systematic uncertainty in the
connection of UV flux density to an actual star formation rate, and
our working assumption $\lambda_{\rm UV}=1400$ \AA\hs does not affect
our main conclusions at all.

For the star formation calibrators discussed above, the ratio
SFR(\lya{})/SFR(UV) is
\begin{equation}
  \frac{{\rm SFR}({\rm Ly}\alpha)}{{\rm SFR (UV)}}=\frac{9.1\times
    10^{-43}L_{{\rm Ly}\alpha}}{1.4\times 10^{-28}L_{\nu,UV}}
  \label{eq:sfr}
\end{equation}
The \lya{} rest-frame equivalent width is defined as

\begin{equation}
  {\rm REW}\equiv \frac{L_{{\rm Ly}\alpha}}{L_{\lambda,{\rm
	Ly}\alpha}}=\frac{\lambda_{{\rm Ly}\alpha}L_{{\rm
	Ly}\alpha}}{\nu_{{\rm Ly}\alpha}L_{\nu,{\rm Ly}\alpha}},
  \label{eq:ew}
\end{equation}
with $L_{\lambda,{\rm Ly}\alpha}$ ($L_{\nu,{\rm Ly}\alpha}$) the flux
density in ergs\,s$^{-1}$\,\AA$^{-1}$ (ergs\,s$^{-1}$\,Hz$^{-1}$) of
the continuum just redward of the \lya{} emission line, and
$\lambda_{{\rm Ly}\alpha}=1216$\,\AA{} ($\nu_{{\rm
Ly}\alpha}=2.46\times 10^{15}$\,Hz) denotes the wavelength (frequency)
of the \lya{} transition. We further used the identity $\lambda
L_{\lambda}=\nu L_{\nu}$ to obtain the right hand side of
Eq~\ref{eq:ew}.

We combine Eq~\ref{eq:sfr} and Eq~\ref{eq:ew} and find
\begin{equation}
  \frac{{\rm SFR}({\rm Ly}\alpha)}{{\rm SFR (UV)}}=\frac{{\rm
      REW}}{{\rm REW}_c}\frac{L_{\nu,{\rm Ly}\alpha}}{L_{\nu,UV}},
  \label{eq:ratio1}
\end{equation}
with REW$_c\equiv \frac{1.4\times 10^{-28}\lambda_{{\rm
      Ly}\alpha}}{9.1\times 10^{-43}\nu_{{\rm
      Ly}\alpha}}=76$\,\AA{}. The ratio of the continuum flux
      densities at $\lambda=1216$\,\AA{} and $\lambda_{\rm
      UV}=1400$\,\AA{} ($L_{\nu,{\rm Ly}\alpha}/L_{\nu,UV}$) depends
      on the slope of the continuum. This slope is usually denoted by
      the parameter $\beta\equiv -d\log L_{\lambda}/d\log \lambda$
      (i.e. $L_{\lambda} \propto \lambda^{-\beta}$ and $L_{\nu}\propto
      \nu^{\beta-2}$). We rewrite Eq~\ref{eq:ratio1} as
\begin{equation}
  \frac{{\rm SFR}({\rm Ly}\alpha)}{{\rm SFR (UV)}}=\Big{(}\frac{{\rm
      REW}}{{\rm REW}_c}\Big{)}\Big{(}\frac{\nu_{{\rm
      Ly}\alpha}}{\nu_{\rm UV}}\Big{)}^{\beta-2}\equiv
      C\Big{(}\frac{{\rm REW}}{{\rm REW}_c}\Big{)}.
      \label{eq:ratio2}
\end{equation}
Eq~\ref{eq:ratio2} shows that the ratio ${\rm SFR}({\rm
Ly}\alpha)/{\rm SFR (UV)}$ is determined uniquely by REW and $\beta$,
because REW$_c$, $\nu_{{\rm Ly}\alpha}$, and $\nu_{\rm UV}$ are
constants. The precise choice\footnote{For example, another conversion
factor that is often  found in the literature is SFR(UV)$=1.25\times
10^{-28}L_{\nu,UV}$ $M_{\odot}$ yr$^{-1}$
\citep[][]{Madau98}. Additionally, the conversion from \lya{}
luminosity to SFR depends on gas metallicity \citep[see][]{S03}.} of
star formation calibrator enters entirely through the value of
REW$_c$, and thus the slope of the correlation. Provided that the
same star formation calibrators are applied consistently to an
ensemble of  galaxies, the actual star formation rates in these
galaxies are irrelevant to the existence of the correlation.

Furthermore, {\it any scatter in the correlation between REW and ${\rm
SFR}({\rm Ly}\alpha)/{\rm SFR (UV)}$ enters entirely through scatter
in $\beta$}.  \lya{} emitting galaxies at $z \sim 5.7$ show that $0
\la \beta \la 2.4$ for $>90$\,\% of the galaxies
\citep{Tapken07}. This is consistent with the median value $\beta_{\rm
med} \sim 1.4$ for \lya{} emitting objects at $z \sim 3$ determined by
e.g. \citet{V05}. Thus, we conservatively adopt $\beta=1.2\pm 1.2$,
which translates to $C=0.89^{+0.16}_{-0.14}$ for $\lambda_{\rm
UV}=1400$\,\AA.

\begin{figure*}
  \plottwo{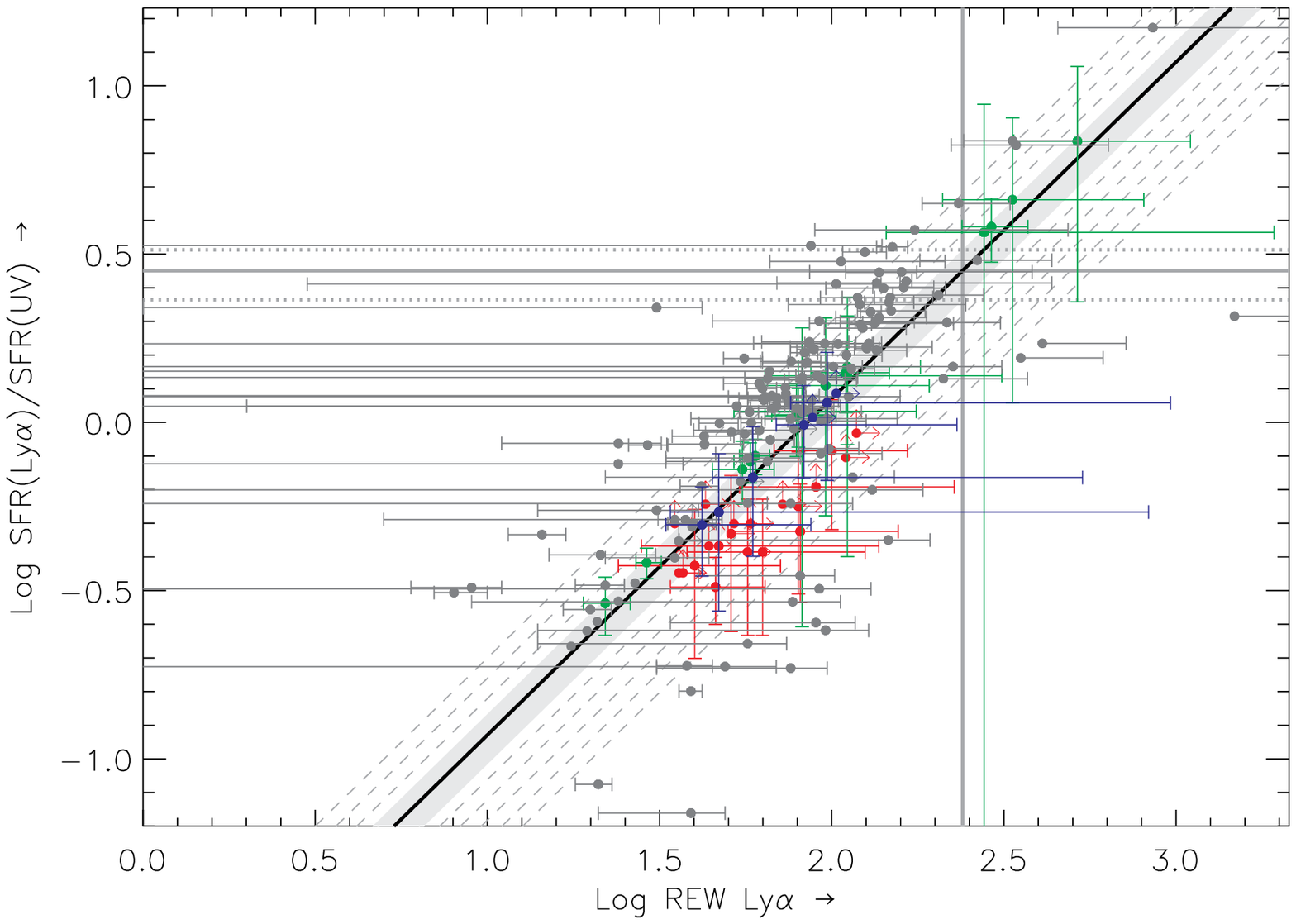}{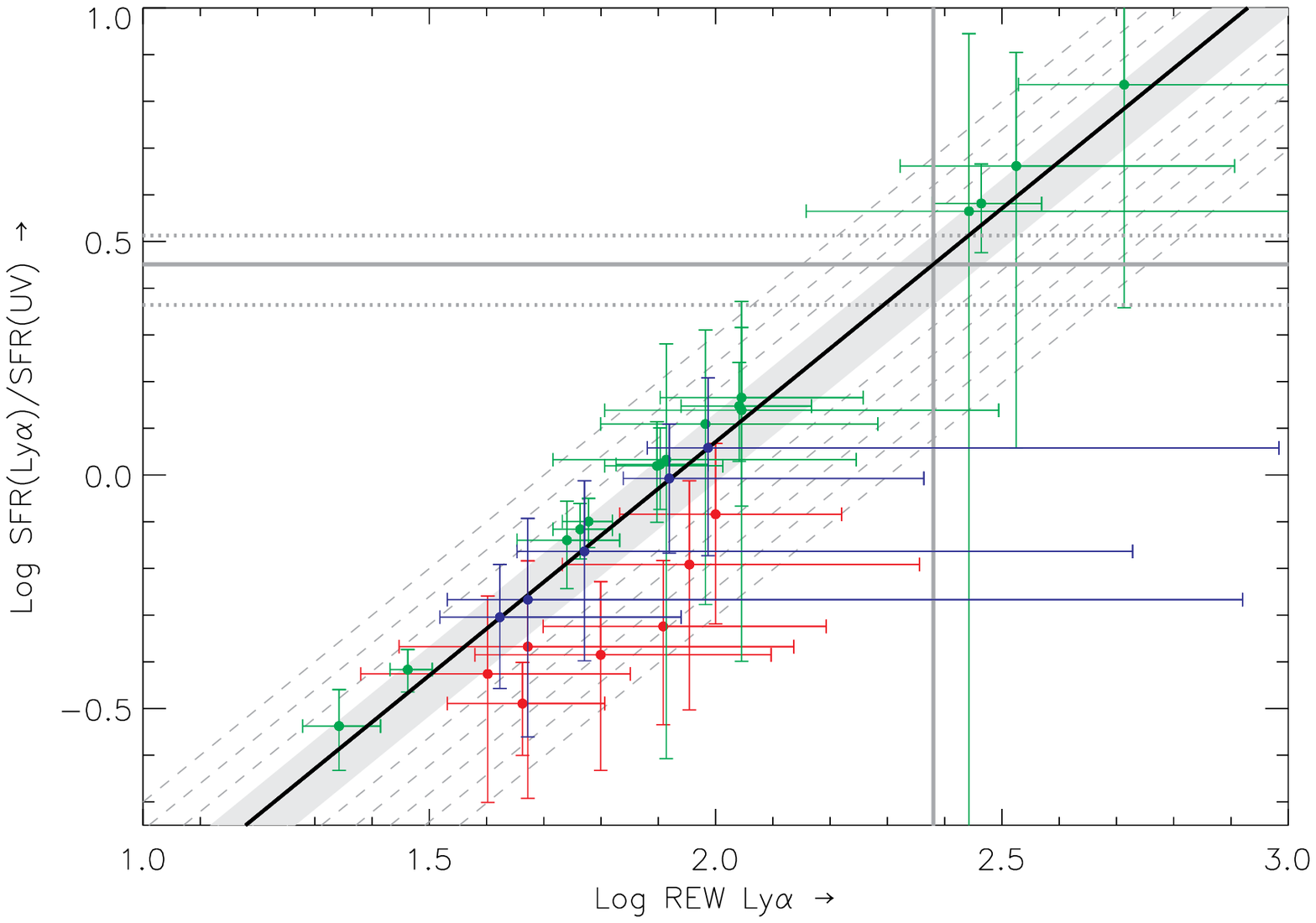}
  \caption{Comparison of the predicted correlation
    (Eq~\ref{eq:ratio2}, {\it solid line}) and its 1\,$\sigma$
    uncertainties ({\it grey region}) with published data (see
    text). Errorbars on all data points denote 1\,$\sigma$
    uncertainties. Where errorbars are missing, the paper contained
    insufficient information to compute them. We show the relation
    SFR(\lya{})/SFR(UV) = $C$ (REW/REW$_c$) as the {\it black solid
    line}, where the {\it grey shaded region} denotes the 1\,$\sigma$
    uncertainty in $C$. The {\it grey dashed lines} indicate the
    locations of galaxies that would have spectra with slopes $\beta =
    [-5, 5]$ with steps of 1. The vertical {\it thick solid} line
    denotes the maximum REW$_{\rm max}=240$\,\AA{} that can be emitted
    by `normal' star forming galaxies
    \citep[e.g.][]{MR02}. Eq~\ref{eq:ratio2} converts this maximum REW
    into a maximum ratio SFR(\lya{})/SFR(UV), which is indicated with
    the horizontal {\it thick solid grey line}, and its 1\,$\sigma$
    uncertainties are indicated with the horizontal {\it thick dotted
    lines}. The majority of data is consistent with the correlation
    given by Eq.~\ref{eq:ratio2}. Statistically significant outliers
    correspond to galaxies with an unusual spectral slope of their
    continuum flux density, $\beta$. {\it Right Panel:} Same as {\it
    left panel} but focusing on the data points with published
    uncertainties on both REW and SFR(Ly$\alpha$)/SFR(UV).}
  \label{fig:corr1}
\end{figure*}

\subsection{Processes that affect the Correlation}
\label{sec:prac}

Obtaining $L_{{\rm Ly}\alpha}$ and $L_{\nu,UV}$ from the measured
quantities $F_\mathrm{Ly\alpha}$ (the \lya{} flux in erg
\,s$^{-1}$\,cm$^{-2}$), and $f_{\nu,UV}$ (the UV continuum flux
density in erg \,s$^{-1}$\,cm$^{-2}$\,Hz$^{-1}$) is non-trivial as it
requires various ``corrections'' that account for the fact that only a
fraction of all emitted UV and \lya{} photons are observed. None
of these corrections are trivial:
\begin{enumerate}
\item For example, the IGM is expected to be opaque (transparent)
  to photons that were emitted blueward (redward) of the \lya{}
  resonance. To first order, this implies that only $\sim 50\%$ of the
  \lya{} photons are transmitted to the observer. However, peculiar
  velocities of intergalactic gas \citep[e.g.][]{igm,Iliev08}, and
  radiative transfer effects in the ISM of galaxies
  \citep[e.g.][]{Ahn03,V08} make this correction quite uncertain. The
  escape fraction of \lya{} photons from the ISM of galaxies is
  heavily regulated by the presence and distribution of dust, as well
  as HI gas kinematics \citep{Kunth98,Hayes,Ostlin,Atek}.  This can
  cause the \lya{} escape fraction to vary significantly between
  objects, or even between different  sight-lines within the same
  object \citep{Laursen}.

\item Dust also affects the UV continuum. The spectral slope in the UV
continuum correlates with the amount of dust extinction
\citep[e.g.][]{Ca94,Heckman98}. The UV continuum slope for unobscured
star forming galaxies, $\beta_{\rm int}$, varies between $0-2.6$
depending on the age, initial mass function, and the star formation
history of a galaxy \citep[see e.g Figs~31 and 32 of][]{LH95}. Dust
lowers the observed UV continuum slope to $\beta=\beta_{\rm
int}-\Delta \beta$, in which $\Delta \beta\sim1.0$[E(B-V)/0.3]
(Calzetti et al. 2000; where E(B-V) denotes the colour excess).

Dust reduces the overall UV flux density by a factor of
$\sim$exp(-E(B-V)/0.1) \citep[see e.g.][]{V08}. The measured slope of
the UV continuum $\beta$ and/or the colour excess E(B-V) can constrain
the amount of extinction of the UV continuum by dust. However these
constraints are uncertain and depend on the intrinsic UV continuum
and/or on the precise shape of the extinction curve \citep[see
e.g.][]{V08}.

\end{enumerate}
Combined these effects introduce large uncertainties in the
conversion from observed Ly$\alpha$ flux to intrinsic Ly$\alpha$
luminosity, and from observed UV flux density to intrinsic UV
luminosity density. Of course,  corrections for dust and/or the IGM
should affect the left and right hand side of Eq~\ref{eq:ratio2}
equally. In practise however, the UV continuum at $\lambda\sim
\lambda_{{\rm Ly}\alpha}$ is often measured directly from the
spectrum, while the continuum at $\lambda= \lambda_{\rm UV}$ is
determined from broadband imaging. Similarly, the \lya{} flux that
enters the left and right hand side of Eq~\ref{eq:ratio2} may be
obtained differently, e.g. from either the spectrum or from narrowband
imaging. Clearly, this can introduce different systematic
uncertainties  to the measurements of SFR(\lya{})/SFR(UV) and
EW. These uncertainties relate to, e.g. slit losses when taking
spectra, dependence on aperture size and seeing for imaging, the
precise filter curve of the narrowband filter \citep[see e.g.][for
extended discussions on complications that arise when determining
$\beta$ and REW from broad and narrow band
measurements]{Hayes06,Hayes}.

Despite these complications, the quoted uncertainties on
SFR(\lya{})/SFR(UV) and EW should reflect the systematic
uncertainties, and we generally expect the data points to be
consistent with Eq~\ref{eq:ratio2}. Any statistically significant
deviation implies that ({\it i}) some corrections are not applied
consistently between the two measurements [e.g. the IGM is corrected
for when determining the EW, but not when determining SFR(\lya{})],
({\it ii}) uncertainties in either EW or SFR(\lya{})/SFR(UV) have been
underestimated, or ({\it iii}) the continuum of the galaxy has an
unusual spectral slope.

The first two possibilities suggest that Eq~\ref{eq:ratio2} provides a
convenient sanity check to whether the uncertainties in
SFR(\lya{})/SFR(UV) and EW are estimated properly. The alternative, an
unusual spectral slope is of great scientific interest. For example,
objects that are dominated by nebular emission, such as galaxies that
contain population III stars \citep{S02} or cooling clouds
\citep{D09}, may be dominated by the two-photon continuum at
$1216\,\mathrm{\AA} < \lambda < 1600\,\mathrm{\AA}$ (for which $\beta
\ll 0$).

\subsection{The Correlation in Existing Data}
\label{sec:data}

We show the REW of LAEs versus the ratio of the SFRs for several
surveys in Figure~\ref{fig:corr1}. The redshifts of the surveys
range from $z\sim 3-6.5$. The data are from \citet{Fujita03},
\citet{Taniguchi05}, \citet{Shima06}, \citet{Ouchi08} ({\it grey data
points}), \citet[][{\it red points}]{Ajiki03}, \citet[][{\it green
points}]{Dawson04}, and \citet[][{\it blue points}]{Venemans04}. We
represent the data from \citet{Shima06} with two sets of points. We
derive ({\it i}) the Ly$\alpha$ flux from the narrowband filter, and
the rest-frame UV flux density from the $z$-band, and ({\it ii}), the
Ly$\alpha$ flux from the spectrum, and the rest-frame UV flux density
from the $i$-band. We use either the published values of REW and
SFR(\lya{})/SFR(UV), or -- when these are not readily available --
computed their values from the available data. In theory, the
correlation in Eq.~\ref{eq:ratio2} is independent of
redshift. However, there are too few measurements available to
observationally confirm this. The {\it left panel} of
Figure~\ref{fig:corr1} shows that the vast majority of galaxies are
consistent with the relation given by Eq.~\ref{eq:ratio2} and appear
to be normal star forming galaxies.

The {\it right panel} of Figure~\ref{fig:corr1} highlights the data
points with published uncertainties in both REW and
SFR(Ly$\alpha$)/SFR(UV) from the {\it left panel}. These points
illustrate more clearly that there are no significant outliers from
the correlation given by Eq~\ref{eq:ratio2}. We performed a
least-squares linear fit to these data points of the form
$\log\frac{{\rm SFR(Ly}\alpha{\rm )}}{{\rm SFR(UV)}}=a+b\log\frac{{\rm
REW}}{{\rm REW}_c}$, and obtained $a=-0.07\pm 0.22$, and $b=1.03\pm
0.11$. This corresponds to $\beta\sim 0.8\pm 3.6$. Stronger
constraints on $\beta$ are not yet possible given the uncertainties in
the data.

\section{Discussion and Conclusions}
\label{sec:conc}
We investigate the relation between REW and the ratio of star
formation rates derived from \lya{} flux, and rest-frame UV flux
density [SFR(\lya{})/SFR(UV)] (Eq~\ref{eq:ratio2}). This relation
derives directly from the definition of equivalent width and star
formation rate conversion factors, and its only source of scatter is
the variation in the slope of the UV continuum at $\lambda_{{\rm
Ly}\alpha}< \lambda <\lambda_{\rm UV}$ between individual
galaxies. The correlation exists regardless of the assumed star
formation calibrators (which themselves depend on the assumed IMF and
gas metallicity), or the true star formation rates of these galaxies.

Despite their fundamentally tight relation, \lya{} REW and
SFR(\lya{})/SFR(UV) are often discussed as independent quantities. We
investigate their correlation in existing data, and find the vast
majority of galaxies to be consistent with the predicted relation (see
Figs~\ref{fig:corr1}). The existence of the relation has
interesting applications, which are discussed next.

\subsection{An Empirically Constrained Ly$\alpha$ Based Star Formation Indicator}

A SFR derived from the UV flux density is likely more reliable than a
SFR derived from Ly$\alpha$ flux. Ly$\alpha$ scatters through the ISM
and IGM which makes it hard to determine the amount of extinction. Our
relation can be used to derive a more accurate Ly$\alpha$ based star
formation calibrator, if we require that --statistically-- the
Ly$\alpha$ derived SFR should be equal to the UV derived SFR. We
introduce the constant $\mathcal{M}$ such that

\begin{equation}
{\rm SFR}_{\mathcal{M}}({\rm Ly}\alpha)\equiv \mathcal{M}\times {\rm
SFR(Ly}\alpha{\rm )}\equiv {\rm SFR(UV)}.
\label{eq:m}
\end{equation}
The constant $\mathcal{M}$ ensures that the SFR derived for a certain
galaxy from its measured Ly$\alpha$ flux is equal to that derived from
its UV flux density. According to Eq~\ref{eq:ratio2}
$\mathcal{M}=[C({\rm REW}/{\rm REW}_{\rm c})]^{-1}$; this implies that
a galaxy with an unusually large REW has $\mathcal{M}\ll 1$. Without
this correction one would overestimate the SFR from the Ly$\alpha$
flux alone.

In the most general case, the probability $P(\mathcal{M})d\mathcal{M}$
that $\mathcal{M}$ lies in the range $\mathcal{M}\pm d\mathcal{M}/2$
is given by (see Appendix~\ref{app:m})

\begin{eqnarray}
P(\mathcal{M})d\mathcal{M}=d\mathcal{M}\hs\mathcal{N}\int_{-\infty}^{\infty}
P(\beta) P({\rm REW}_{\mathcal M})\frac{{\rm REW}_{\mathcal
M}(\beta)}{\mathcal{M}}d\beta,
\label{eq:pm}
\end{eqnarray}
with $\mathcal{N}$ the normalization factor which ensures that
$\int_{0}^{\infty} P(\mathcal{M})d\mathcal{M}=1$, and
REW$_{\mathcal{M}}(\beta)\equiv {\rm REW}_{\rm
c}/(\mathcal{M}C)$. Furthermore, $P(\mathrm{REW})d\mathrm{REW}$
denotes the probability that a LAE has an observed REW in the range
REW$\pm d$REW/2; $P(\beta)d\beta$ denotes the probability that a LAE
has an observed $\beta$ in the range $\beta \pm d\beta/2$.

\begin{figure}
\vbox{\centerline{\epsfig{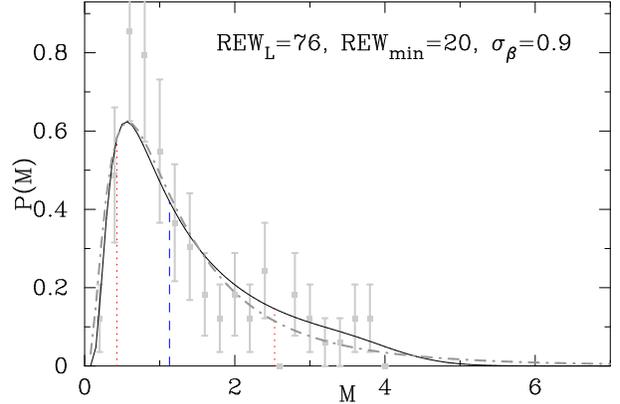}}}
\caption[]{The {\it black solid line} shows the probability
distribution for $\mathcal{M}\equiv$SFR(UV)/SFR(Ly$\alpha$) as given
by Eq~\ref{eq:ratio2}, in which $P$(REW) and $P$($\beta$) are
determined from the observations (see text). The {\it red dotted
lines} mark the $68\%$ confidence interval centered on the median
$\mathcal{M}=1.13$ (indicated by the {\it blue dashed line}). We thus
find that$(68\%$ of LAEs have $\mathcal{M}=1.1^{+1.4}_{-0.7}$. The
data points denote the directly measured distribution of
SFR(UV)/SFR(Ly$\alpha$) collected from the papers used in our
analysis. The errorbars denote Poisson uncertainties in these
data. The observed data distribution agrees well with our derived
PDF. Furthermore, the derived PDF is reproduced quite well by a
log-normal distribution which is indicated by the {\it grey dot-dashed
line}, which is easily adopted when modeling statistical properties of
LAEs.}
\label{fig:pdf}
\end{figure}

Figure~\ref{fig:pdf} shows the probability distribution
$P(\mathcal{M})$ obtained from Eq~\ref{eq:pm} ({\it black solid
line}). In this calculation we assume: ({\it i}) $P(\beta)d\beta$ is a
Gaussian with $\bar{\beta}=1.2$, and $\sigma_{\beta}=0.9$. This choice
ensures that $\sim 90\%$ of the galaxies have $0 \la \beta \la 2.4$
\citep[cf.][]{Tapken07}; ({\it ii}) $P(\mathrm{REW})d\mathrm{REW}$ is
an exponential with a scale length of REW$_{\rm L}$=76
\AA\hs\citep[][]{Gronwall07}, based on observed Ly$\alpha$ emitting
galaxies at $z=3.1$ with REW$> 20$ \AA, i.e $P$(REW)dREW$\propto$
exp(-REW/REW$_{\rm L})$ for REW$>$REW$_{\rm min}\sim$ 20 \AA, and
P(REW)$=0$ otherwise. In Appendix~\ref{app:dos} we show that the
precise shape of the distribution is not very sensitive to the assumed
probability density functions (PDFs) of $\beta$ and REW.

Figure~\ref{fig:pdf} shows that $P(\mathcal{M})$ peaks at
$\mathcal{M}\sim 0.6$. The function $P(\mathcal{M})$ is highly
asymmetric and its median is $\mathcal{M}=1.13$, which is indicated as
the {\it blue dashed line}. That is, SFR(Ly$\alpha$) $<0.88$ SFR(UV)
for $50\%$ of the galaxies (similarly, SFR(Ly$\alpha$) $<$ SFR(UV) for
$\sim 57\%$ of the galaxies). The {\it red dotted lines} mark the
$68\%$ confidence interval centered on the median value. We therefore
find that for $68\%$ of LAEs $\mathcal{M}=1.1^{+1.4}_{-0.7}$. In other
words, the Ly$\alpha$ derived SFR lies within a factor of $\sim 2.5$
from the UV derived SFR for 68\% of LAEs. Overplotted as the data
points is the measured distribution of SFR(UV)/SFR(Ly$\alpha$)
collected from the papers that were used in our analysis. The
errorbars denote Poisson uncertainties. This observed distribution
agrees quite well with our derived PDF. The {\it grey dot-dashed line}
indicates a log-normal distribution
\begin{eqnarray}
P(\mathcal{M})d\mathcal{M}=\frac{1}{\sigma\sqrt{2\pi}}{\rm exp}\Big{[}
  -\frac{1}{2}\Big{(}\frac{[\log
      \mathcal{M}]-x}{\sigma}\Big{)}^2\Big{]}\frac{d\mathcal{M}}{\mathcal{M}\ln
  10},
\label{eq:log}
\end{eqnarray}
with $x=0.04$ and $\sigma=0.35$. This log-normal distribution provides
a decent fit to the derived and observed distribution, and is easily
adopted when modeling statistical properties of LAEs.

\citet{Ouchi08} conclude that between $z=3$ and $z=6$ the observed REW
PDF is consistent with no redshift evolution. This implies that our
derived PDF for $\mathcal{M}$ also applies at redshifts greater than
$z=3$. On the other hand, the measured scalelength of the exponential
REW distribution at $z=2.3$ is REW$_{\rm L}=48$
\AA\hs\citep{Nilsson09}, causing the PDF to broaden and shift to
larger $\mathcal{M}$ (see Appendix
\ref{app:dos}). Table~\ref{table:param} summarizes the redshift
dependence of the parameters describing the log-normal distribution
for $\mathcal{M}$. In both redshift bins the standard deviation  of
the log-normal distribution is very similar, with $\sigma \sim
0.3-0.35$.
The abrupt change of the value of $x$ at $z=3$ is clearly a crude
approximation of the real redshift evolution of the
$\mathcal{M}$-PDF. We have chosen this parametrization because it
corresponds to the simplest description that is consistent with
existing data.

\begin{table}
\begin{minipage}{8cm}
\centering
\caption{Fit Parameters for the log-normal PDF for $P(\log
\mathcal{M})d\mathcal{M}$.}
\begin{tabular}{l c c}
\hline\hline & x & $\sigma$\\ $z \gsim 3$\hs\footnote{We assume that
 the REW-PDF, and hence the $\mathcal{M}$-PDF, does not to evolve
 between $z=3-6$ (Ouchi et al. 2008). We take the observed $z=3$
 REW-PDF from Gronwall et al. (2007) for $z\gsim 3$, and the observed
 $z=2.3$ REW-PDF from Nilsson et al. (2009) for $z\lsim3$.} & 0.04 &
 0.35 \\ $z \lsim 3$ & 0.20& 0.31 \\ \hline\hline
\end{tabular}
\label{table:param}
\end{minipage}
\end{table}

In theoretical work the inverse problem often arises in which a
Ly$\alpha$ luminosity must be obtained from a physical SFR. In such a
case one may write $L_{{\rm Ly}\alpha}=\frac{1.1\times 10^{42}\hs{\rm
erg}\hs{\rm s}^{-1}}{\mathcal{M}}\Big{(}\frac{{\rm
SFR}}{M_{\odot}\hs{\rm yr}^{-1}}\Big{)}$ and adopt our PDF for
$\mathcal{M}$. This allows one to assign an empirically calibrated,
variable Ly$\alpha$ luminosity to galaxies of a given SFR. This
prescription is clearly not perfect, because not all galaxies that are
actively forming stars show a Ly$\alpha$ emission line. For example,
only 20-25$\%$ of Lyman Break galaxies (LBGs) at $z=3$ have a
Ly$\alpha$ line with REW$\gsim 20\%$ \AA\hs and would classify as a
LAE \citep[e.g.][]{Shapley03}, and the connection between LBGs and
LAEs is not well understood. Nevertheless, our suggested prescription
for assigning Ly$\alpha$ luminosities to galaxies  of a given SFR is
more realistic than the often used one-to-one relation $L_{{\rm
Ly}\alpha}=1.1 \times 10^{42}\hs{\rm erg}\hs{\rm s}^{-1}\hs {\rm
SFR}/(M_{\odot}\hs{\rm yr}^{-1})$.

\subsection{Outliers in the SFR(Ly$\alpha$)/SFR(UV)-REW Plane: Signposts for Unusual Galaxies?}

``Normal'' star forming galaxies can emit a maximum REW$_{\rm
  max}=240$\,\AA. Galaxies with REW$>$REW$_{\rm max}$ may signal the
  presence of a galaxy that contains population III stars
  \citep[e.g.][]{MR02}. Using Eq~\ref{eq:ratio2} this corresponds to
  [SFR(\lya{})/SFR(UV)]$_{\rm max} =2.8^{+0.4}_{-0.5}$. The majority
  of observed galaxies that have quoted uncertainties on
  SFR(\lya{})/SFR(UV) are consistent with this upper limit.

Objects where nebular emission dominates, such as galaxies containing
population III stars \citep{S02} or cooling clouds \citep{D09}, may be
dominated by the two-photon continuum at 1216\,\AA{} $< \lambda <$
1600\,\AA{}. This results in unusually negative values for
$\beta$. These objects may have been identified in the spectrum
itself, because deviations from the correlation are caused by unusual
spectral slopes. However, reliable measurements of the continuum just
redward of the Ly$\alpha$ line and at $\lambda_{\rm UV}=1400$ \AA\hs
provide a long baseline in wavelength, which may more clearly reveal
the presence of a continuum dominated by two-photon emission. This
suggests that outliers in the REW - SFR(Ly$\alpha$)/SFR(UV) plane may
provide a more sensitive probe to cooling clouds or primordial
galaxies than the spectrum alone. That additional probe is important
especially at high redshifts, where the IGM may transmit only a small
fraction of the \lya{} emitted by galaxies \citep{igm}. In this case
even those star forming galaxies containing population III stars may
have REW$<$REW$_{\rm max}$. This strongly suggests that the determined
REW alone is not enough to identify a population III galaxy.

{\bf Acknowledgements} M.D. is supported by Harvard University
funds. E.W. acknowledges the Smithsonian Institution for the support
of his postdoctoral fellowship. We thank an anonymous referee for
helpful, constructive comments that improved the content of this paper.

\appendix
\section{The Probability Distribution Function (PDF) for $\mathcal{M}$}
\subsection{Derivation of $P(\mathcal{M})d\mathcal{M}$}
\label{app:m}
We employ the notation of probability theory, where the function
$p(y|b)$ denotes the conditional probability density function (PDF) of
$y$ given $b$. The PDF for $y$ is then given by $p(y)=\int
p(y|b)p(b)db$, where $p(b)$ denotes the PDF for $b$. Similarly, we can
write $p(y)=\int \int p(y|b,a)p(b)p(a)dbda$, where $p(y|b,a)$ denotes
the conditional probability density function (PDF) of $y$ given $b$
and $a$, and where we assume that $a$ and $b$ are independent. We can
thus write the probability $P(\mathcal{M})d\mathcal{M}$ that
$\mathcal{M}$ lies in the range $\mathcal{M}\pm d\mathcal{M}/2$ as

\begin{eqnarray}
P(\mathcal{M})d\mathcal{M}=d\mathcal{M}\hs
\mathcal{N}\int_{-\infty}^{\infty}d\beta\int_{{\rm EW}_{\rm
min}}^{\infty}d{\rm REW}\hs \nonumber \\ P(\mathcal{M}|\beta,{\rm
REW}) P({\rm REW})P(\beta),
\label{eq:spm}
\end{eqnarray}
with REW$_{\rm min}$ the minimum REW, and $P(\mathcal{M}|\beta,{\rm
REW})$ the conditional PDF $\mathcal{M}$ given\footnote{We assume that
REW and $\beta$ are independent variables. In the case that larger
data sets demonstrate that this assumption is false, then one needs to
replace $ P({\rm REW})P(\beta)$ with $P({\rm REW}|\beta)P(\beta)$ in
Eq~\ref{eq:spm}.} $\beta$ and REW.

Because we know that $\int_0^{\infty}P(\mathcal{M}|\beta,{\rm
REW})d\mathcal{M}=1$, and that for a given $\beta$ and REW there is
only one solution $\mathcal{M}=[C({\rm REW}/{\rm REW}_{\rm c})]^{-1}$
(Eq~\ref{eq:m}), we write $P(\mathcal{M}|\beta,{\rm REW})=\delta_{\rm
D}(\mathcal{M}-[C({\rm REW}/{\rm REW}_{\rm c})]^{-1})\equiv
\delta_{\rm D}(g[{\rm REW}])$, where $\delta_D(x)$ denotes the
standard Dirac-delta function. We eliminate the integral over REW by
using the property of Dirac delta functions that $\delta_{\rm
D}(h[x])=\frac{\delta(x-x_0)}{|h'(x_0)|}$, where $x_0$ denotes the
(real) root of $h(x)$. If we denote the root of $g({\rm REW})$ with
REW$_{\mathcal{M}}$ then $|g'({\rm
REW}_{\mathcal{M}})|=\frac{\mathcal{M}}{{\rm REW}_{\mathcal{M}}}$ and
obtain Eq~\ref{eq:pm}.

\subsection{Dependence of $P(\mathcal{M})d\mathcal{M}$ on Model Parameters}
\label{app:dos} 

\begin{figure}
\vbox{\centerline{\epsfig{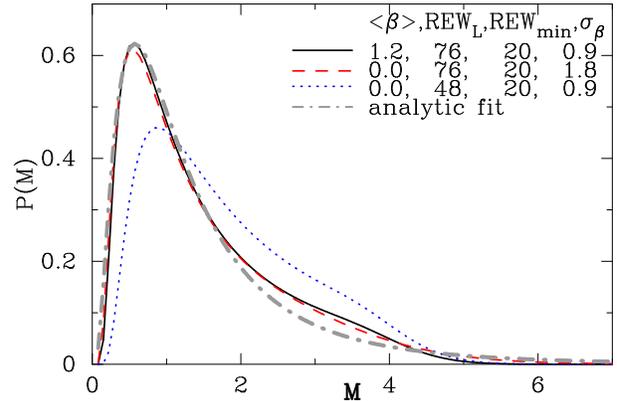}}}
\caption[]{This Figure shows the dependence of
$P(\mathcal{M})d\mathcal{M}$ on various model parameters that describe
the PDFs for $\beta$ and REW, which include $\bar{\beta}$,
$\sigma_{\beta}$, REW$_{\rm min}$ and REW$_{\rm L}$. Each curve
represent a calculation in which one of the model parameters - as
indicated in the top right corner of the Figure - was varied.}
\label{fig:pdf2}
\end{figure}
\begin{figure}
\vbox{\centerline{\epsfig{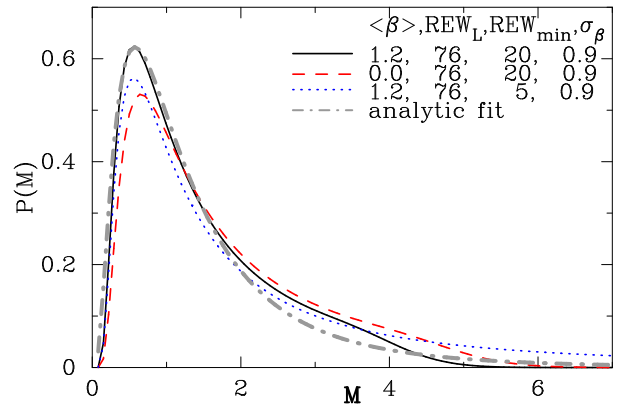}}}
\caption[]{Same as Fig~\ref{fig:pdf2} but for model parameters
indicated in the top right corner.}
\label{fig:pdf3}
\end{figure}
We investigate the dependence of $P(\mathcal{M})d\mathcal{M}$ on
various model parameters that describe the PDFs for $\beta$ and REW,
which include $\bar{\beta}$, $\sigma_{\beta}$, REW$_{\rm min}$ and
REW$_{\rm L}$. Figures~\ref{fig:pdf2} and \ref{fig:pdf3} show the PDFs
that we obtain when varying one of the model parameters as indicated
in the top right corner of each Figure. For example, the {\it red
dashed line} shows the PDF that we obtain for $\bar{\beta}=0.0$.

Figures~\ref{fig:pdf2} and \ref{fig:pdf3} show that the PDF does not
vary significantly when changing the parameters $\bar{\beta}$,
$\sigma_{\beta}$, REW$_{\rm min}$. The PDF appears to be most
sensitive to the scale length of exponential REW distribution. The
value REW$_{\rm L}=48$ \AA\hs corresponds to the scale length that was
derived by \citet{Nilsson09} for $z=2.3$ LAEs. At these lower
redshifts there is a lower fraction of large EW emitters, which pushes
the $\mathcal{M}$ PDF to larger values.

\label{lastpage}
\end{document}